\documentstyle[prl,aps]{revtex}

\begin{document}
\author{Jian-Qi Shen $^{1,}$$^{2}$\footnote{E-mail address: jqshen@coer.zju.edu.cn}}
\address{ $^{1}$ Centre for Optical
and Electromagnetic Research, State Key Laboratory of Modern
Optical Instrumentation, Zhejiang University, Hangzhou Yuquan
310027, P. R. China\\
 $^{2}$ Zhejiang Institute of Modern Physics and
Department of Physics, Zhejiang University, Hangzhou 310027, P. R.
China}
\date{\today }
\title{Frequency-independent effective rest mass of photons \\in the 2TDLM model}
\maketitle

\begin{abstract}
A physically interesting {\it effective rest mass} of photons in
electromagnetic media, which is independent of wave frequency
$\omega$, is defined in the present paper. It is verified that
this frequency-independent effective rest mass of photons can be
easily read off from the optical refractive index squared
$n^{2}\left( \omega \right) $ of commonly-seen electromagnetic
media. As an illustrative example, we extract the
frequency-independent effective rest mass of photons from
$n^{2}\left( \omega \right) $ in the {\it two time derivative
Lorentz material} (2TDLM) model. The connection between effective
rest mass and electromagnetic parameters of electric permittivity
($\epsilon $) and magnetic permeability ($\mu $) in left-handed
media is also briefly discussed.

\end{abstract}
\section{Introduction}
The effective mass of electrons and holes in a semiconductor
possesses the physical significance: it contains details and
information on the interaction between electrons and lattices, and
provides us with a vivid and convenient description of motion of
electrons and electron-lattice couplings in a solid state.
Moreover, this effective mass can be measured experimentally. So,
the effective mass of electrons in a semiconductor is regarded as
an essential concept\cite{Burns}, which is helpful for describing
the electron wave propagating inside the semiconductor (and the
solid state). Likewise, it is reasonably believed that the {\it
effective rest mass} of photons in media may also be physically
meaningful. In the present paper, two kinds of effective rest mass
({\it i.e.}, the frequency-dependent and frequency-independent
rest mass) of photons will be suggested. The definition of {\it
frequency-dependent} effective rest mass of photons in media will
be merely briefly discussed and its deficiencies will also be
demonstrated from the physical point of view. To eliminate the
drawback of the frequency-dependent effective rest mass, we will
suggest a new definition of effective rest mass, {\it i.e.}, the
{\it frequency-independent} photon rest mass in electromagnetic
media and consider it in more detail. It is shown that the
frequency-independent effective rest mass of photons can be
extracted from the optical refractive index squared $n^{2}\left(
\omega \right) $. As an illustrative example, we obtain the
expression for the frequency-independent effective rest mass in
the {\it two time derivative Lorentz material} (2TDLM)
model\cite{Ziolkowski2,Ziolkowski3,Ziolkowski4}.

Recently, metamaterials such as left-handed media\cite{Veselago},
has captured attention of a large number of researchers in various
fields, {\it e.g.}, materials science, condensed matter physics,
optics and electromagnetism as
well\cite{Ziolkowski2,Ziolkowski3,Ziolkowski4}. One of the
metamaterials models proposed by Ziolkowski and Auzanneau is
called the {\it two time derivative Lorentz material} (2TDLM)
model\cite{Ziolkowski3,Ziolkowski4}, which is a generalization of
the standard Lorentz material model and encompasses the
permittivity and permeability material responses experimentally
obtained, which may be seen, for example, in the reference of
Smith {\it et al.}\cite{Smith}.

This paper is organized as follows: first we consider the
frequency-dependent effective rest mass of photons in media. The
formulation of the frequency-independent effective rest mass is
presented in Sec. III. In Sec. IV, the frequency-independent
effective rest mass in the 2TDLM model and left-handed media is
read off from $n^{2}\left( \omega \right) $. In Sec. V, we
conclude with some remarks on the frequency-independent effective
rest mass of photons in media.

\section{Frequency-dependent effective rest mass of photons in media}

We have studied the wave propagation problem ({\it i.e.},
polarization of plane wave, geometric phases, time evolution of
photon wavefunction, helicity reversal) of photons field in media
(optical fiber) at the quantum level\cite{Shen,Zhu}, where we have
not considered the dispersion and absorption properties of media.
In the present paper, these electromagnetic and optical properties
of media contributing to the wave propagation of light field will
be taken into account. It is readily verified that the interaction
of light wave with media can give rise to an {\it effective rest
mass} of photons, which may be illustrated as follows:

if the concept of {\it effective rest mass} is applicable to the
photons in media, then in terms of the Einstein-de Broglie
relation, a photon possessing $m_{\rm eff}$ agrees with the
dispersion relation $\left( \hbar k\right) ^{2}c^{2}=\left( \hbar
\omega \right) ^{2}-m_{\rm eff}^{2}c^{4}$ with $\hbar $ and $c$
being the Planck's constant and the speed of light in a vacuum,
respectively. Thus one can arrive at $ k^{2}c^{2}=\omega
^{2}\left( 1-\frac{c^{4}}{\hbar ^{2}}\frac{m_{\rm eff}^{2}}{\omega
^{2}}\right)$ or
\begin{equation}
\frac{\omega ^{2}}{k^{2}}=\frac{c^{2}}{1-\frac{c^{4}}{\hbar
^{2}}\frac{m_{\rm eff}^{2}}{\omega ^{2}}}.         \label{eq21}
\end{equation}
Since the relation between the frequency $\omega $ and wave vector
$k$ of light wave propagating inside a dispersive medium may be
$\frac{\omega ^{2}}{k^{2}}=\frac{c^{2}}{n^{2}}$ with $n$ being the
optical refractive index. So, with the help of (\ref{eq21}) and
dispersion relation $\frac{\omega
^{2}}{k^{2}}=\frac{c^{2}}{n^{2}}$, one can conclude that the
effective rest mass $m_{\rm eff}$, refractive index $n$ and wave
frequency $\omega $ together satisfy the following relation

\begin{equation}
\frac{m_{\rm eff}^{2}c^{4}}{\hbar ^{2}}=(1-n^{2})\omega ^{2}.
                 \label{eq102}
\end{equation}
Thus we obtained the expression for the effective rest mass
squared $m_{\rm eff}^{2}$. For Lorentz dispersive media, the
refractive index squared is of the form\cite{Fang}
\begin{equation}
n^{2}=\epsilon=1+\frac{Ne^{2}}{\epsilon _{0} m_{\rm
e}}\frac{1}{\omega _{0}^{2}-\omega ^{2}-i\gamma \omega },
\end{equation}
and the effective rest mass squared is therefore written
\begin{equation}
m_{\rm eff}^{2}=\frac{\hbar ^{2}}{c^{2}}\left( \frac{\mu
_{0}Ne^{2}}{m_{\rm e}}\frac{\omega ^{2}}{\omega ^{2}+i\gamma
\omega -\omega _{0}^{2}}\right),  \label{eq24}
\end{equation}
where use is made of the relation $c^{2}=\frac{1}{\epsilon _{0}\mu
_{0}}$. Here $\gamma $ and $\omega _{0}$ respectively denote the
damping frequency and resonance frequency of the electric dipole
oscillators; $N$ and $m_{\rm e}$ represent the electron number per
unit volume and electron mass, respectively. Note that this
effective rest mass of photons in electromagnetic media originates
from the interaction term $\mu _{0}{\bf J}\cdot {\bf A}$ in the
Lagrangian density, where $\bf J$ and ${\bf A}$ respectively
denote the electric current density and the magnetic vector
potentials. Since the electric current acted upon by external
electromagnetic wave behaves like the self-induced charge
current\cite{Ho}, $\bf J$ is proportional to ${\bf A}$, for
instance, in Lorentz materials, the electric current density in
the presence of electromagnetic wave is given
\begin{equation}
{\bf J}=Ne{\bf v}=\frac{Ne^{2}\omega ^{2}}{m_{\rm
e}}\frac{1}{\omega _{0}^{2}-\omega ^{2}-i\gamma \omega }{\bf A}.
\end{equation}
Thus the interaction term $\mu _{0}{\bf J}\cdot {\bf A}$ is
changed into a mass term $-\frac{m_{\rm eff}^{2}c^{2}}{\hbar
^{2}}{\bf A}^{2}$ of electromagnetic fields in the Lagrangian
density ( see in the Appendix for the detailed discussions on this
problem ). Apparently, if $\gamma =0$ and $\omega _{0}^{2}=0$,
then the expression (\ref {eq24}) for $m_{\rm eff}^{2}$ is reduced
to the rest mass squared that is independent of $\omega $. We are
certain that in this case ($\gamma =\omega _{0}=0$), $m_{\rm eff}$
in (\ref{eq24}) possesses a form of plasma-type effective rest
mass $m_{\rm eff}=\frac{\hbar \omega _{\rm p}}{c^{2}}$ with
$\omega _{\rm p}^{2}=\frac{\mu _{0}c^{2}Ne^{2}}{m_{\rm
e}}$\cite{Feinberg}. It is readily verified that $L=\frac{\hbar
}{m_{\rm eff}c}$ is just the Londons' penetration length ({\it
i.e.}, the magnetic field can penetrate only the scale $L$ within
the surface of the superconductor). Thus it is concluded without
fear that the above formulation for effective rest mass of photons
in media is self-consistent.

However, it should be noted that in Lorentz dispersive materials,
$m_{\rm eff}^{2}$ in (\ref{eq24}) depends strongly on $\omega$,
and sometimes possesses an imaginary part (due to the damping
parameter and resonance frequency) and the physical meanings of
$m_{\rm eff}^{2}$ is therefore not very apparently seen. In what
follows we will define a physically meaningful effective rest mass
of photons, which is independent of the frequency $\omega$, and
will explain a method by which the frequency-independent effective
rest mass of photons in the 2TDLM media is extracted from
$n^{2}\left( \omega \right) $. Note that the above non-physical
disadvantage of frequency-dependent mass in (\ref{eq24}) will be
avoided in the frequency-independent effective rest mass.

\section{Extracting frequency-independent effective mass from refractive index squared}
In this framework of definition of frequency-independent effective
rest mass, the photon should be regarded as acted upon by a
hypothetical force field, namely, it behaves like a massive de
Broglie particle moving in a force field. It is verified in the
following that, for the case of de Broglie particle in a force
field, one can also consider its ``optical refractive index" $n$.
According to the Einstein- de Broglie relation, the dispersion
relation of the de Broglie particle with rest mass $m_{0}$ in the
potential field $V\left( {\bf x}\right) $ agrees with
\begin{equation}
\left( \omega -\phi \right)
^{2}=k^{2}c^{2}+\frac{m_{0}^{2}c^{4}}{\hbar ^{2}} \label{eqA7}
\end{equation}
with $\phi =\frac{V}{\hbar }$. It follows that
\begin{equation}
\omega ^{2}\left[ 1-\frac{m_{0}^{2}c^{4}}{\hbar ^{2}\omega
^{2}}-2\frac{\phi }{\omega }+\left( \frac{\phi }{\omega }\right)
^{2}\right] =k^{2}c^{2},     \label{eqA8}
\end{equation}
which yields
\begin{equation}
\frac{\omega
^{2}}{k^{2}}=\frac{c^{2}}{1-\frac{m_{0}^{2}c^{4}}{\hbar ^{2}\omega
^{2}}-2\frac{\phi }{\omega }+\left( \frac{\phi }{\omega }\right)
^{2}}.                                        \label{eqA9}
\end{equation}
Compared with the dispersion relation $\frac{\omega
^{2}}{k^{2}}=\frac{c^{2}}{n^{2}}$, one can arrive at
\begin{equation}
n^{2}=1-\frac{m_{0}^{2}c^{4}}{\hbar ^{2}\omega ^{2}}-2\frac{\phi
}{\omega }+\left( \frac{\phi }{\omega }\right) ^{2},
                                   \label{eq34}
\end{equation}
which is the square of ``optical refractive index" of the de
Broglie particle in the presence of a potential field $\phi $.

If a photon is permitted to possess a frequency-independent
effective rest mass $m_{\rm eff}$, then the expression
(\ref{eq34}) for the case of photons in media is rewritten
\begin{equation}
n^{2}\left( \omega\right) =1-\frac{m_{\rm eff}^{2}c^{4}}{\hbar
^{2}\omega ^{2}}-2\frac{\phi }{\omega }+\left( \frac{\phi }{\omega
}\right) ^{2}.            \label{eq35}
\end{equation}
In what follows, an approach to frequency-independent effective
mass extracted from the optical refractive index squared
$n^{2}\left( \omega\right) $ is presented. It is assumed that
$n^{2}\left( \omega\right) $ can be rewritten as the following
series expansions
\begin{equation}
n^{2}\left( \omega \right) =\sum_{k}\frac{a_{-k}}{\omega ^{k}},
\end{equation}
then one can arrive at $n^{2}\left( \omega \right) \omega
^{2}=\sum_{k}\frac{a_{-k}}{\omega ^{k-2}}$, namely,
\begin{equation}
n^{2}\left( \omega \right) \omega ^{2}= ...+\frac{a_{-3}}{\omega
}+a_{-2}+a_{-1}\omega +a_{0}\omega ^{2}+a_{+1}\omega ^{3}+...\quad
. \label{eq37}
\end{equation}
Compared (\ref{eq37}) with (\ref{eq35}), it is apparent that the
frequency-independent effective rest mass squared $m_{\rm
eff}^{2}$ of the photon can be read off from $n^{2}\left( \omega
\right)$, namely, the constant term on the right-handed side on
(\ref{eq37}) is related only to $m_{\rm eff}^{2}$, {\it i.e.},
\begin{equation}
a_{-2}=-\frac{m_{\rm eff}^{2}c^{4}}{\hbar ^{2}}.
\end{equation}

Generally speaking, the optical refractive index $n^{2}\left(
\omega \right) $ is often of complicated form, particularly for
the metamaterials. So, we cannot extract frequency-independent
$m_{\rm eff}^{2}$ immediately from $n^{2}\left( \omega \right) $.
In the following, this problem is resolved by calculating the
limit value of $n^{2}\left( \omega \right) \omega ^{2}$, where
$\omega $ tends to $\infty $ or zero. We first consider the large
frequency behavior of $n^{2}\left( \omega \right) \omega ^{2}$. It
follows from (\ref{eq37}) that $n^{2}\left( \omega \right) \omega
^{2}$ at high frequencies ({\it i.e.}, $\omega \rightarrow \infty
$) is of the form
\begin{equation}
n^{2}\left( \omega \right) \omega ^{2}{\rightarrow
}a_{-2}+a_{-1}\omega +a_{0}\omega ^{2}+a_{+1}\omega
^{3}+a_{+2}\omega ^{4}+...\quad .
\end{equation}
This, therefore, means that $n^{2}\left( \omega \right) \omega
^{2}$ at high frequencies ($\omega \rightarrow \infty $) can be
split into the following two terms
\begin{equation}
n^{2}\left( \omega \right) \omega ^{2}{\rightarrow }\left( {\rm
constant \quad term}\right) +\left( {\rm divergent \quad
term}\right).
\end{equation}
So, we can obtain $m_{\rm eff}^{2}$ from the following formula
\begin{equation}
\lim_{\omega \rightarrow \infty }\left[ {n^{2}\left( \omega
\right) \omega ^{2}-\left( {\rm divergent \quad
term}\right)}\right]=-\frac{m_{\rm eff}^{2}c^{4}}{\hbar ^{2}}.
\label{eq311}
\end{equation}
Note that the constant term in $n^{2}\left( \omega \right) \omega
^{2}$ can also be extracted by considering the low-frequency
(zero-frequency) behavior of $n^{2}\left( \omega \right) $. It is
certain that in this case the obtained constant term is the same
as (\ref{eq311}). Apparently $n^{2}\left( \omega \right) \omega
^{2}$ at low-frequencies ({\it i.e.}, $\omega \rightarrow 0$)
behaves like
\begin{equation}
n^{2}\left( \omega \right) \omega ^{2}\rightarrow
...+\frac{a_{-5}}{\omega ^{3}}+\frac{a_{-4}}{\omega
^{2}}+\frac{a_{-3}}{\omega }+a_{-2}, \label{eq312}
\end{equation}
where the sum term ($...+\frac{a_{-5}}{\omega
^{3}}+\frac{a_{-4}}{\omega ^{2}}+\frac{a_{-3}}{\omega }$) is
divergent if $\omega$ approaches zero. Thus, subtraction of the
divergent term from $n^{2}\left( \omega \right) \omega ^{2}$
yields
\begin{equation}
\lim_{\omega \rightarrow 0}\left[ n^{2}\left( \omega \right)
\omega ^{2}-\left( {\rm divergent \quad term}\right) \right]
=-\frac{m_{\rm eff}^{2}c^{4}}{\hbar ^{2}}.       \label{eq313}
\end{equation}

In the next section, we will calculate the frequency-independent
effective rest mass of photons in 2TDLM model by making use of the
method presented above.
\section{Frequency-independent effective rest mass of photons in 2TDLM model}
According to the definition of the 2TDLM model, the
frequency-domain electric and magnetic susceptibilities are
given\cite{Ziolkowski5}
\begin{equation}
\chi ^{e}\left( \omega \right)=\frac{\left( \omega _{\rm
p}^{e}\right) ^{2}\chi _{\alpha }^{e}+i\omega \omega _{\rm
p}^{e}\chi _{\beta }^{e}-\omega ^{2}\chi _{\gamma }^{e}}{-\omega
^{2}+i\omega \Gamma ^{e}+\left( \omega _{0}^{e}\right) ^{2}},
\quad   \chi ^{m}\left( \omega \right)=\frac{\left( \omega _{\rm
p}^{m}\right) ^{2}\chi _{\alpha }^{m}+i\omega \omega _{\rm
p}^{m}\chi _{\beta }^{m}-\omega ^{2}\chi _{\gamma }^{m}}{-\omega
^{2}+i\omega \Gamma ^{m}+\left( \omega _{0}^{m}\right) ^{2}},
\label{eq41}
\end{equation}
where $\chi _{\alpha }^{e,m}$, $\chi _{\beta }^{e,m}$ and $\chi
_{\gamma }^{e,m}$ represent, respectively, the coupling of the
electric (magnetic) field and its first and second time
derivatives to the local electric (magnetic) dipole motions.
$\omega _{\rm p}$, $\Gamma ^{e}$, $\Gamma ^{m}$ and $\omega _{0}$
can be viewed as the plasma frequency, damping frequency and
resonance frequency of the electric (magnetic) dipole oscillators,
respectively. So, the refractive index squared in the 2TDLM model
reads
\begin{equation}
n^{2}\left( \omega \right) =1+\chi ^{e}\left( \omega \right) +\chi
^{m}\left( \omega \right) +\chi ^{e}\left( \omega \right) \chi
^{m}\left( \omega \right).                \label{eq42}
\end{equation}
In order to obtain $m_{\rm eff}^{2}$ from $n^{2}\left( \omega
\right)$, according to the formulation in the previous section,
the high-frequency ($\omega \rightarrow \infty $) behavior of
$\chi ^{e}\left( \omega \right) \omega ^{2}$ should first be taken
into consideration and the result is
\begin{equation}
\chi ^{e}\left( \omega \right) \omega ^{2}\rightarrow -i\omega
\omega _{\rm p}^{e}\chi _{\beta }^{e}+\omega ^{2}\chi _{\gamma
}^{e}.                   \label{eq43}
\end{equation}
Note that $-i\omega \omega _{\rm p}^{e}\chi _{\beta }^{e}+\omega
^{2}\chi _{\gamma }^{e}$ is the divergent term of $\chi ^{e}\left(
\omega \right) \omega ^{2}$ at large frequencies. Subtracting the
divergent term from $\chi ^{e}\left( \omega \right) \omega ^{2}$,
we consider again the high-frequency behavior ($\omega \rightarrow
\infty $) and obtain another divergent term $i\omega ^{3}\Gamma
^{e}\chi _{\gamma }^{e}$, {\it i.e.},
\begin{equation}
\chi ^{e}\left( \omega \right) \omega ^{2}-\left( -i\omega \omega
_{\rm p}^{e}\chi _{\beta }^{e}+\omega ^{2}\chi _{\gamma
}^{e}\right) =\frac{\omega ^{2}\left( \omega _{\rm p}^{e}\right)
^{2}\chi _{\alpha }^{e}-\omega ^{2}\omega _{\rm p}^{e}\Gamma
^{e}\chi _{\beta }^{e}+i\omega \omega _{\rm p}^{e}\left( \omega
_{0}^{e}\right) ^{2}\chi _{\beta }^{e}-i\omega ^{3}\Gamma ^{e}\chi
_{\gamma }^{e}-\omega ^{2}\left( \omega _{0}^{e}\right) ^{2}\chi
_{\gamma }^{e}}{-\omega ^{2}+i\omega \Gamma ^{e}+\left( \omega
_{0}^{e}\right) ^{2}}\rightarrow i\omega ^{3}\Gamma ^{e}\chi
_{\gamma }^{e}.                 \label{eq44}
\end{equation}
Thus, after subtracting all the divergent terms from $\chi
^{e}\left( \omega \right) \omega ^{2}$, we obtain
\begin{equation}
\lim_{\omega \rightarrow \infty }\left[ \chi ^{e}\left( \omega
\right) \omega ^{2}-\left( -i\omega \omega _{\rm p}^{e}\chi
_{\beta }^{e}+\omega ^{2}\chi _{\gamma }^{e}\right) -i\omega
^{3}\Gamma ^{e}\chi _{\gamma }^{e}\right] =-\left[ \left( \omega
_{\rm p}^{e}\right) ^{2}\chi _{\alpha }^{e}-\omega _{\rm
p}^{e}\Gamma ^{e}\chi _{\beta }^{e}-\left( \omega _{0}^{e}\right)
^{2}\chi _{\gamma }^{e}+\left( \Gamma ^{e}\right) ^{2}\chi
_{\gamma }^{e}\right].                           \label{eq45}
\end{equation}
In the same fashion, we obtain
\begin{equation}
\lim_{\omega \rightarrow \infty }\left[ \chi ^{m}\left( \omega
\right) \omega ^{2}-\left( -i\omega \omega _{\rm p}^{m}\chi
_{\beta }^{m}+\omega ^{2}\chi _{\gamma }^{m}\right) -i\omega
^{3}\Gamma ^{m}\chi _{\gamma }^{m}\right] =-\left[ \left( \omega
_{\rm p}^{m}\right) ^{2}\chi _{\alpha }^{m}-\omega _{\rm
p}^{m}\Gamma ^{m}\chi _{\beta }^{m}-\left( \omega _{0}^{m}\right)
^{2}\chi _{\gamma }^{m}+\left( \Gamma ^{m}\right) ^{2}\chi
_{\gamma }^{m}\right]                      \label{eq46}
\end{equation}
for the magnetic susceptibility $\chi ^{m}\left( \omega \right) $.
In what follows we continue to consider the term $\chi ^{e}\left(
\omega \right) \chi ^{m}\left( \omega \right) $ on the
right-handed side of (\ref{eq42}). In the similar manner, the
infinite-frequency limit of $\chi ^{e}\left( \omega \right) \chi
^{m}\left( \omega \right) \omega ^{2}$ is given as follows
\begin{eqnarray}
\lim_{\omega \rightarrow \infty }\chi ^{e}\left( \omega \right)
\chi ^{m}\left( \omega \right) \omega ^{2}&=&-\chi _{\gamma
}^{m}\left[ \left( \omega _{\rm p}^{e}\right) ^{2}\chi _{\alpha
}^{e}-\omega _{\rm p}^{e}\Gamma ^{e}\chi _{\beta }^{e}-\left(
\omega _{0}^{e}\right) ^{2}\chi _{\gamma }^{e}+\left( \Gamma
^{e}\right)
^{2}\chi _{\gamma }^{e}\right]                               \nonumber \\
&-&\chi _{\gamma }^{e}\left[ \left( \omega _{\rm p}^{m}\right)
^{2}\chi _{\alpha }^{m}-\omega _{\rm p}^{m}\Gamma ^{m}\chi _{\beta
}^{m}-\left( \omega _{0}^{m}\right) ^{2}\chi _{\gamma }^{m}+\left(
\Gamma ^{m}\right) ^{2}\chi _{\gamma }^{m}\right].    \label{eq47}
\end{eqnarray}
Hence, insertion of (\ref{eq45})-(\ref{eq47}) into (\ref{eq311})
yields
\begin{eqnarray}
\frac{m_{\rm eff}^{2}c^{4}}{\hbar ^{2}}&=&\left( 1+\chi _{\gamma
}^{m}\right) \left[ \left( \omega _{\rm p}^{e}\right) ^{2}\chi
_{\alpha }^{e}-\omega _{\rm p}^{e}\Gamma ^{e}\chi _{\beta
}^{e}-\left( \omega _{0}^{e}\right) ^{2}\chi _{\gamma }^{e}+\left(
\Gamma ^{e}\right) ^{2}\chi _{\gamma
}^{e}\right]                \nonumber \\
&+&\left( 1+\chi _{\gamma }^{e}\right)\left[ \left(\omega _{\rm
p}^{m}\right) ^{2}\chi _{\alpha }^{m}-\omega _{\rm p}^{m}\Gamma
^{m}\chi _{\beta }^{m}-\left( \omega _{0}^{m}\right) ^{2}\chi
_{\gamma }^{m}+\left( \Gamma ^{m}\right) ^{2}\chi _{\gamma
}^{m}\right].                \label{eq48}
\end{eqnarray}
Thus the $\omega$-independent effective rest mass squared $m_{\rm
eff}^{2}$ is extracted from $n^{2}\left( \omega \right)$.
Apparently, $m_{\rm eff}^{2}$ in Eq. (\ref{eq48}) does not depend
on $\omega$. The remainder on the right-handed side of Eq.
(\ref{eq35}) is $n^{2}\left( \omega \right) -1+\frac{m_{\rm
eff}^{2}c^{4}}{\hbar ^{2}\omega ^{2}}$, which is equal to $\left(
\frac{\phi }{\omega }\right) ^{2}-2\frac{\phi }{\omega }$. For
simplicity, we set
\begin{equation}
\Delta \left( \omega \right) =n^{2}\left( \omega \right)
-1+\frac{m_{\rm eff}^{2}c^{4}}{\hbar ^{2}\omega ^{2}},
\end{equation}
and then from the equation
\begin{equation}
\left( \frac{\phi }{\omega }\right) ^{2}-2\frac{\phi }{\omega
}=\Delta \left( \omega \right),
\end{equation}
one can readily obtain the expression for the hypothetical
potential field $\phi \left( \omega \right)$ is $\phi \left(
\omega \right) =\omega \left[ 1- \sqrt{1+\Delta \left( \omega
\right) }\right]$, or
\begin{equation}
V\left( \omega \right) =\hbar \omega \left[ 1- \sqrt{1+\Delta
\left( \omega \right) }\right],  \label{eq411}
\end{equation}
where $\phi =\frac{V}{\hbar }$ has been previously defined in Sec.
III.

Many metamaterials which have complicated expressions for electric
and magnetic susceptibilities such as (\ref{eq41}) are generally
rarely seen in nature and often arises from the artificial
manufactures. More recently, one of the artificial composite
metamaterials, the left-handed medium which has a frequency band
(GHz) where the electric permittivity ($\epsilon $) and the
magnetic permeability ($\mu $) are simultaneously negative, has
focused attention of many authors both experimentally and
theoretically\cite{Ziolkowski2,Smith,Klimov,Pendry2,Pendry3,Shelby}.
In the left-handed medium, most phenomena as the Doppler effect,
Vavilov-Cherenkov radiation and even Snell's law are inverted. In
1964\footnote{Note that, in the literature, many authors mention
the year when Veselago suggested the {\it left-handed media} by
mistake. They claim that Veselago proposed the concept of {\it
left-handed media} in 1968. On the contrary, the true fact is as
follows: Veselago's excellent paper was first published in July,
1964 [Usp. Fiz. Nauk {\bf 92} 517-526]. In 1968, this original
paper was translated into English by W. H. Furry and published
again in the journal of Sov. Phys. Usp.\cite{Veselago}.}, Veselago
first considered many peculiar optical and electromagnetic
properties, phenomena and effects in this medium and referred to
such materials as left-handed media\cite{Veselago}, since in this
case the propagation vector $\bf k$, electric field $\bf E$ and
magnetic field $\bf H$ of light wave propagating inside it form a
left-handed system. It follows from Maxwellian curl equations that
such media having negative simultaneously negative $\epsilon $ and
$\mu $ exhibit a negative index of refraction, {\it i.e.},
$n=-\sqrt{\epsilon \mu }$. Since negative refractive index occurs
only rarely, this medium attracts attention of many physicists in
various fields. In experiments, the negative $\epsilon $ and $\mu
$ can be respectively realized by using a network (array) of thin
(long) metal wires\cite{Pendry2} and a periodic arrangement of
split ring resonators\cite{Pendry3}. A combination of the two
structures yields a left-handed medium. In general, the dielectric
parameter $ \epsilon \left( \omega \right)$ and magnetic
permeability $\mu \left( \omega \right)$ in the isotropic
homogeneous left-handed medium are of the form
\begin{equation}
 \epsilon \left( \omega \right)
=1-\frac{\omega _{\rm p}^{2}}{\omega \left( \omega +i\gamma
\right) },\quad \mu \left( \omega \right) =1-\frac{F\omega
^{2}}{\omega ^{2}-\omega _{0}^{2}+i\omega \Gamma }
\label{eq412}
\end{equation}
with the plasma frequency $\omega _{\rm p}$ and the magnetic
resonance frequency $\omega _{0}$ being in the GHz region.
$\gamma$ and $\Gamma$ stand for the electric and magnetic damping
parameters, respectively. Compared (\ref{eq412}) with the electric
and magnetic susceptibilities (\ref{eq41}), one can arrive at
\begin{eqnarray}
\chi _{\alpha }^{e}=1, \quad \omega _{\rm p}^{e}=\omega _{\rm p},
\quad \chi _{\beta }^{e}=\chi _{\gamma }^{e}=0, \quad \omega
_{0}^{e}=0, \quad  \Gamma ^{e}=-\gamma
,                                         \nonumber \\
 \chi _{\gamma }^{m}=-F, \quad \chi _{\alpha }^{m}=\chi _{\beta
}^{m}=0, \quad \omega _{0}^{m}=\omega _{0}, \quad \Gamma
^{m}=-\Gamma.    \label{eq413}
\end{eqnarray}
So, it is readily verified with the help of (\ref{eq48}) that the
$\omega$-independent effective rest mass of photons inside
left-handed media is written as follows
\begin{equation}
\frac{m_{\rm eff}^{2}c^{4}}{\hbar ^{2}}=\left( 1-F\right) \omega
_{\rm p}^{2}+F\left(\omega _{0}^{2}-\Gamma ^{2}\right).
\label{eq414}
\end{equation}

It follows that Eq. (\ref{eq414}) is a restriction on the
electromagnetic parameters $F, \omega _{\rm p},  \omega _{0},
\Gamma $ and $F$ in the electric permittivity and magnetic
permeability (\ref{eq412}), since $m_{\rm eff}^{2} \geq 0$.
Insertion of the experimentally chosen values of the
electromagnetic parameters in the literature into Eq.
(\ref{eq414}) shows that this restriction condition is satisfied.
For instance, in Ruppin's work\cite{Ruppin}, $F, \omega _{\rm p},
\omega _{0}, \gamma, \Gamma $ and $F$ were chosen $F=0.56, \quad
\omega _{\rm p}=10.0 {\rm GHz}, \quad \omega _{0}=4.0 {\rm GHz},
\quad \gamma=0.03\omega _{\rm p},\quad \Gamma=0.03\omega _{0}$,
which indicates that the inequality $m_{\rm eff}^{2} \geq 0$ is
satisfied.

\section{Concluding remarks}
This paper deals mainly with a problem of how to extract the
frequency-independent effective rest mass of photons in media from
the optical refractive index squared $n^{2}\left( \omega \right)$.
It is of physical interest that the expressions (\ref{eq48}) and
(\ref{eq414}) for $m_{\rm eff}^{2}$ are the mathematical
restrictions on the electromagnetic parameters in the electric
permittivity and magnetic permeability of artificial composite
metamaterials. Apparently, it is certain that the two essential
requirements (\ref{eq48}) and (\ref{eq414}) are related close to
the Kramers-Kronig dispersion relation.

To close this paper, we consider briefly the potential physical
significance of $m_{\rm eff}^{2}$ in electromagnetic media. Since
it is independent of frequency of light wave, this effective rest
mass of photons possesses the explicit physical meanings. It
follows from (\ref{eq48}) that all the parameters in $\epsilon
\left( \omega \right) $ and $\mu \left( \omega \right) $
contribute to $m_{\rm eff}^{2}$ and that $m_{\rm eff}^{2}$
therefore consists of various interactions between light fields
and electromagnetic materials. If $m_{\rm eff}^{2}$ can be
obtained by experimental measurements, then we are able to
investigate the parameters in $\epsilon \left( \omega \right) $
and $\mu \left( \omega \right) $ as well as the electromagnetic
interaction inside media in more detail. It should be noted that,
for the plasma-type media such as the superconductor and the array
of long metal wires (which possesses a negative permittivity in
the GHz range), the above two effective rest mass ({\it i.e.},
frequency-dependent and frequency-independent rest mass) are
equivalent, namely, the frequency-dependent effective rest mass is
reduced to the frequency-independent effective rest mass. In the
former case of effective rest mass (frequency-dependent), the
photon in media is regarded as a free massive de Broglie particle.
But, its effective rest mass lacks explicit physical meanings
since it is dependent strongly on the frequency $\omega $. Whereas
in the latter case of effective rest mass (frequency-independent),
the photon in media may be considered a massive de Broglie
particle acted upon by a hypothetical potential field. Hence, the
frequency-independent effective rest mass has somewhat explicit
physical meanings and may therefore have potential applications to
the investigation of light propagation in electromagnetic media
whose permittivity and permeability are of the complicated form.
We hold that both theoretical and experimental analysis of this
frequency-independent effective rest mass of photons in
metamaterials deserve further considerations.
\\ \\
\textbf{Acknowledgements}  I thank Sai-Ling He for useful
discussions on the wave propagation in left-handed media. This
work is supported in part by the National Natural Science
Foundation of China under the project No. $90101024$.
\\ \\
{\bf Appendix}

In this Appendix, we introduce the {\it effective rest mass} of
electromagnetic wave from the fundamental principle ( based on the
Lagrangian density of electrodynamics ). Note that in Londons'
electromagnetics for superconductivity, Ginzberg-Landau
superconductivity theory and Higgs mechanism, there are also the
viewpoints of {\it effective rest mass}. Here we consider the
effective rest mass of electromagnetic wave in superconducting
media ( or electron plasma ) and dispersive/absorptive materials.
It is well known that the Lagrangian density of electromagnetics
is
\begin{equation}
\ell =-\frac{1}{4}F_{\mu \nu }F^{\mu \nu }-\frac{1}{\epsilon
_{0}c^{2}}\rho \varphi +\mu _{0}{\bf J}\cdot {\bf A}.   \eqnum{A1}
\label{eqB1}
\end{equation}
The equation of motion of an electron in superconducting media (
or electron plasma ) acted upon by the external electromagnetic
wave is $m_{\rm e}\dot{{\bf v}}=e{\bf E}$ with ${\bf
E}=-\frac{\partial }{\partial t}{\bf A}$, where dot on the
velocity ${\bf v}$ denotes the derivative of ${\bf v}$ with
respect to time $t$ and ${\bf A}$ stands for the magnetic vector
potentials of the applied electromagnetic wave. One can therefore
arrive at

\begin{equation}
\frac{{\rm d}}{{\rm d}t}\left( m_{\rm e}{\bf v}+e{\bf A}\right)
=0, \eqnum{A2} \label{eqB2}
\end{equation}
namely, the canonical momentum $m_{\rm e}{\bf v}+e{\bf A}$ of the
electron is conserved. Set $m_{\rm e}{\bf v}+e{\bf A}=0$ and then
$m_{\rm e}{\bf v}=-e{\bf A}$. So, the electric current density is
of the form
\begin{equation}
{\bf J}=Ne{\bf v}=-\frac{Ne^{2}}{m_{\rm e}}{\bf A}   \eqnum{A3}
\label{eqB3}
\end{equation}
with $N$ being the electron number in per unit volume. It follows
from (\ref{eqB1}) that the interaction term $\mu _{0}{\bf J}\cdot
{\bf A}$ in Lagrangian density is rewritten as follows
\begin{equation}
\mu _{0}{\bf J}\cdot {\bf A}=-\frac{\mu _{0}Ne^{2}}{m_{\rm e}}{\bf
A}^{2}=-\frac{Ne^{2}}{\epsilon _{0}m_{\rm e}c^{2}}{\bf A}^{2},
\eqnum{A4} \label{eqB4}
\end{equation}
where use is made of the relation $c^{2}=\frac{1}{\epsilon _{0}\mu
_{0}}$. It is apparently seen that since electron current acts as
the self-induced charge current\cite{Ho}, the interaction term
$\mu _{0}{\bf J}\cdot {\bf A}$ is therefore transformed into the
mass term $-\frac{m_{\rm eff}^{2}c^{2}}{\hbar ^{2}}{\bf A}^{2}$ of
electromagnetic fields. Hence, we have
\begin{equation}
\frac{m_{\rm eff}^{2}c^{2}}{\hbar ^{2}}=\frac{Ne^{2}}{\epsilon
_{0}m_{\rm e}c^{2}}   \eqnum{A5} \label{eqB5}
\end{equation}
and the {\it effective mass} squared
\begin{equation}
m_{\rm eff}^{2}=\frac{\hbar ^{2}}{c^{2}}\left(
\frac{Ne^{2}}{\epsilon _{0}m_{\rm e}c^{2}}\right).   \eqnum{A6}
\label{eqB6}
\end{equation}
It is readily verified that $L=\frac{\hbar }{m_{\rm eff}c}$ is
just the Londons' penetration length ( {\it i. e.}, the magnetic
field can penetrate only the scale $L$ within the surface of the
superconductor ).

In what follows we take into account the case of dispersive and
absorptive materials. Here we only consider the motion of
electrons, and the result presented in the following holds also
for ions in media. The equation of motion of electrons acted upon
by the time harmonic electromagnetic wave in these electromagnetic
materials is written

\begin{equation}
\ddot{{\bf x}}+\gamma \dot{{\bf x}}+\omega _{0}^{2}{\bf
x}=\frac{e}{m_{\rm e}}{\bf E}_{0}\exp \left[ -i\omega t\right].
\eqnum{A7} \label{eqB7}
\end{equation}
Here ${\bf x}$, $\gamma $ and $\omega _{0}$ denote the electron
spatial displacement vector, damping frequency and resonance
frequency of the electric dipole oscillators, respectively. One
can obtains
\begin{equation}
{\bf x}=\frac{e}{m_{\rm e}}\frac{1}{\omega _{0}^{2}-\omega
^{2}-i\gamma \omega }{\bf E}_{0}\exp \left[ -i\omega t\right]
\eqnum{A8} \label{eqB8}
\end{equation}
from Eq. (\ref {eqB7}). Further calculation yields

\begin{equation}
\ddot{{\bf x}}=-\frac{e\omega ^{2}}{m_{\rm e}}\frac{1}{\omega
_{0}^{2}-\omega ^{2}-i\gamma \omega }{\bf E}_{0}\exp \left[
-i\omega t\right] =-\frac{e\omega ^{2}}{m_{\rm e}}\frac{1}{\omega
_{0}^{2}-\omega ^{2}-i\gamma \omega }\left( -\frac{\partial
}{\partial t}{\bf A}\right),   \eqnum{A9} \label{eqB9}
\end{equation}
where use is made of ${\bf E}_{0}\exp \left[ -i\omega t\right]
=-\frac{\partial }{\partial t}{\bf A}$. Both the electron velocity
( ${\bf v}=\dot{{\bf x}}$ ) and the electric current density {\bf
J} are therefore obtained, {\it i. e.},

\begin{equation}
{\bf v}=\frac{e\omega ^{2}}{m_{\rm e}}\frac{1}{\omega
_{0}^{2}-\omega ^{2}-i\gamma \omega }{\bf A}   \eqnum{A10}
\label{eqB10}
\end{equation}
and
\begin{equation}
{\bf J}=Ne{\bf v}=\frac{Ne^{2}\omega ^{2}}{m_{\rm
e}}\frac{1}{\omega _{0}^{2}-\omega ^{2}-i\gamma \omega }{\bf A}.
\eqnum{A11} \label{eqB11}
\end{equation}
Hence, similar to the above procedure (\ref {eqB4})-(\ref {eqB6})
( {\it i. e.}, the effective mass term arises from the interaction
term ), here the effective mass can be obtained from

\begin{equation}
\frac{m_{\rm eff}^{2}c^{2}}{\hbar ^{2}}=-\frac{\mu
_{0}Ne^{2}\omega ^{2}}{m_{\rm e}}\frac{1}{\omega _{0}^{2}-\omega
^{2}-i\gamma \omega } \eqnum{A12} \label{eqB12}
\end{equation}
or
\begin{equation}
m_{\rm eff}^{2}=\frac{\hbar ^{2}}{c^{2}}\left( \frac{\mu
_{0}Ne^{2}}{m_{\rm e}}\frac{\omega ^{2}}{\omega ^{2}+i\gamma
\omega -\omega _{0}^{2}}\right).  \eqnum{A13} \label{eqB13}
\end{equation}
Apparently, if $\gamma =0$ and $\omega _{0}^{2}=0$, then the
expression (\ref {eqB13}) for $m_{\rm eff}^{2}$ is reduced to
(\ref {eqB6}), which is independent of $\omega $. It is certain
that in this case ( $\gamma =\omega _{0}=0$ ), $m_{\rm eff}$ in
(\ref{eqB13}) possesses the form of plasma-type effective mass (
frequency-independent rest mass ).

\end{document}